\documentclass[twocolumn,preprintnumbers,amsmath,amssymb]{revtex4-1}

\usepackage{graphicx}
\usepackage{dcolumn}
\usepackage{bm}

\begin{document}

\title{Bilayer membranes in micro-fluidics: from gel emulsions to soft functional devices}

\author{Shashi Thutupalli}
\email{shashi.thutupalli@ds.mpg.de}
\affiliation{Max Planck Institute for Dynamics and Self-Organization, Bunsenstr. 10, 37073 G\"ottingen, Germany}
\author{Ralf Seemann}
\email{ralf.seemann@ds.mpg.de}
\affiliation{Experimental Physics, Saarland University, Saarbr\"ucken, Germany}
\affiliation{Max Planck Institute for Dynamics and Self-Organization, Bunsenstr. 10, 37073 G\"ottingen, Germany}
\author{Stephan Herminghaus}
\email{stephan.herminghaus@ds.mpg.de}
\affiliation{Max Planck Institute for Dynamics and Self-Organization, Bunsenstr. 10, 37073 G\"ottingen, Germany}

\date{\today}

\begin{abstract}
We outline a concept of self-assembled soft matter devices based on micro-fluidics, which use surfactant bilayer membranes as their main building blocks, arrested in geometric structures provided by top-down lithography. Membranes form spontaneously when suitable water-in-oil emulsions are forced into micro-fluidic channels at high dispersed-phase volume fractions. They turn out to be remarkably stable even when pumped through the micro-fluidic channel system. Their geometric arrangement is self-assembling, driven by interfacial energy and wetting forces. The ordered membrane arrays thus emerging can be loaded with amphiphilic functional molecules, ion channels, or just be used as they are, exploiting their peculiar physical properties. For wet electronic circuitry, the aqueous droplets then serve as the 'solder points'. Furthermore, the membranes can serve as well-controlled coupling media between chemical processes taking place in adjacent droplets. This is shown for the well-known Belousov-Zhabotinski reaction. Suitable channel geometries can be used to (re-) arrange the droplets, and thereby their contents, in a controlled way by just moving the emulsion through the device. It thereby appears feasible to construct complex devices out of molecular-size components in a self-assembled, but well controlled manner.
\end{abstract}

\maketitle

\section{Introduction}

The miniaturization of technical components and machines has been one of the most powerful motors of advancement in both science and technology for the past four decades. Feature sizes in modern electronic circuits have come down to the scale of 45 nanometers, and single transistors are meanwhile routinely made smaller than 100 nm \cite{Benn2007}. As a consequence, the density of components placed on micro-chips has seen a roughly exponential increase over many years. This development, which for its practical robustness has received the colloquial term 'Moore's law', is at the heart of, e.g., the enormous recent increase of widely available computer power \cite{Moore1965}. All this has been reached by means of 'top-down' lithographic techniques, which are capable of structuring solid state materials into arbitrary shapes with amazing accuracy by sophisticated lithographic procedures.

However, it is clear that there will soon be an end to this development of ever smaller structuring. Well before the structures come of molecular size ($\sim 1$ nm), interfacial diffusion will lead to their rapid destruction. Other transport processes directly linked to the function of the device, such as electro-migration in electronic chips, are even much more effective. Since the lithographic approach necessarily leads to a structure which is very far from thermal equilibrium, this must be seen as a fundamental problem to the top-down concept. The large interfacial energy stored in these systems directly drives transport phenomena which are likely to lead to their destruction. Building devices with feature sizes even close to molecular scales therefore appears prohibitively difficult in a top-down approach. Any successful concept for exploiting the full scale down to molecular sizes would have to master the balancing act of stepping far enough out of equilibrium to make sustained dynamical functions possible, but staying close enough to equilibrium to avoid the destructive force of large gradients in free energy.

Yet we know that such systems exist abundantly on earth: all living matter has dynamic functional units on a hierarchy of length scales, down to molecular size. This has been possible only because evolution has used more than three billion years of genotype experience to master the key task posed above on the phenotype level. The building blocks, chiefly lipids, proteins, and nucleic acids, are designed such that they either self-assemble into the desired structure or function, or can be assembled at expense of only small amounts of free energy, such as in the case of chaperonins assembling the 'correct' tertiary structure of proteins \cite{Altschuler2008}. At the same time, chemical energy supply keeps the system sufficiently far off equilibrium to give rise to complex dynamic function, and to maintain structural components as traits of non-equilibrium steady states \cite{BenJac2003}. If we are to build devices with building blocks of molecular size, it therefore appears advisable to follow a similar path exploiting the self-assembly concept \cite{Alivi1999,Lehn2002,Kurth2000,Drain2002}.

It suggests itself to use components similar to those which have been so successfully 'tested' by nature for eons. Soft matter, such as the materials of  living systems, is governed by typical interaction energies on the thermal scale ($\sim kT$), such that non-trivial functions are possible preferentially at room temperature. It exhibits high molecular mobilities as compared to classical solid state materials, like semiconductors, thus enabling sufficiently rapid self-assembly (and re-assembly) processes. Furthermore, the dispersion interactions dominating the structure and dynamics of soft matter systems are small compared to chemical binding energies. As a consequence, all molecules taking actively part in the functional processes stay intact. This opens an almost unlimited variety of building blocks.  Finally, the use of building blocks similar to those of living matter appears particularly promising for devices to be interfaced with living organisms, such as for modern orthotic or prosthetic technology.

However, we cannot rely entirely on self-assembly, since the goal of any design is to reach a certain 'phenotype', the idea of which is at the start of the whole endeavor. Coding the full complexity of the desired device into the molecular building blocks, such that the desired structure emerges completely out of self-assembly, would require full control of non-equilibrium states out of their microscopic conditions; this is a long-standing and so far unsolved problem. Even if a solution was at hand, the task would be still tremendous, and probably impossible to complete. It will thus be necessary to provide a sufficiently strict pre-selection of configurations which are 'allowed' to the system for its assembly. A conceptually straightforward implementation of such pre-selection is some solid scaffold, which provides enough of the desired geometry to subtly 'convince' the soft components to assemble the way one would like them to.

\begin{figure}[h]
\includegraphics[width=8.5cm]{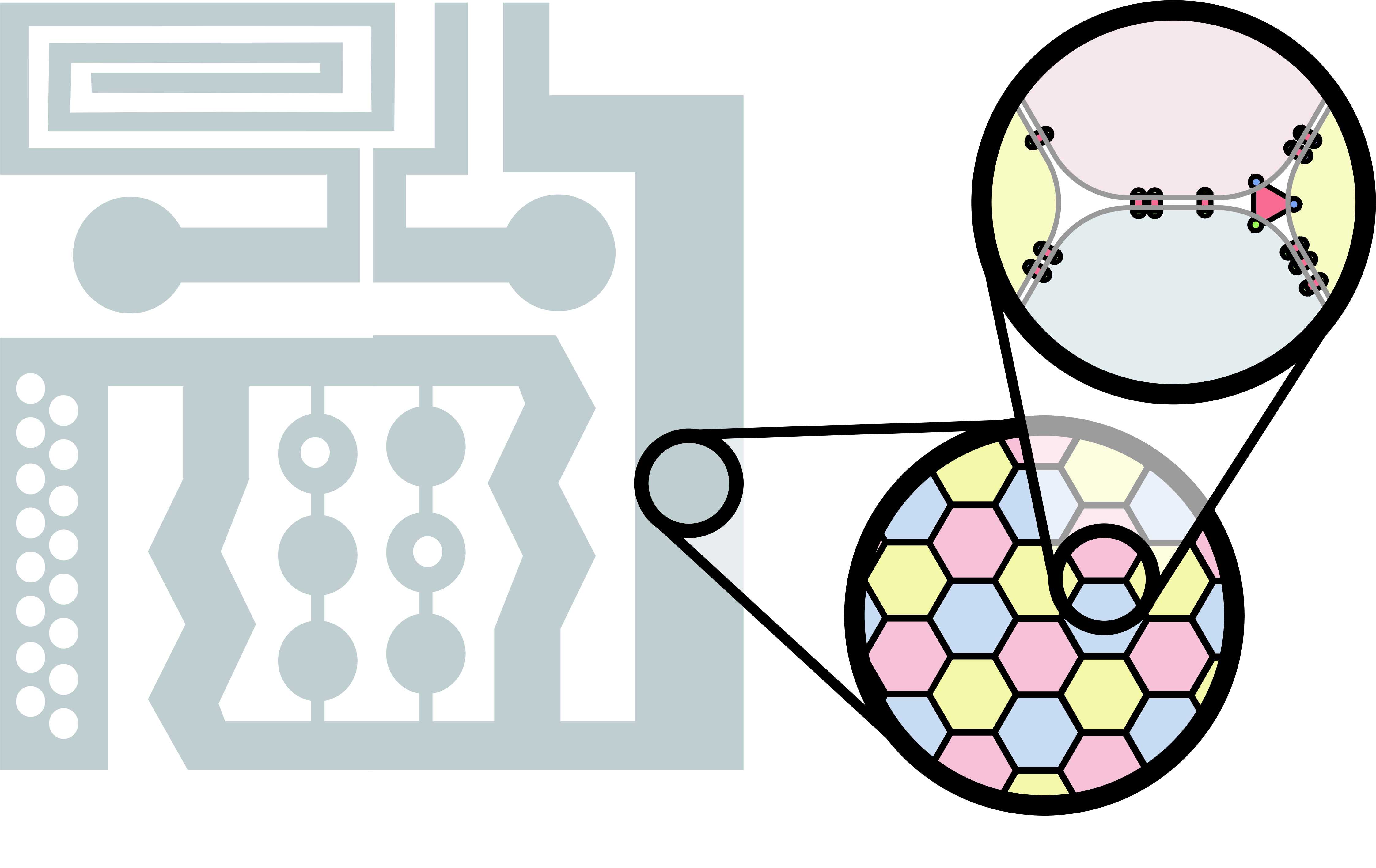}
\caption{Sketch of a concept combining top-down lithographic design of liquid channel structures (left, grey pattern) with complex gel emulsions forming self-assembled droplet and membrane arrangements within these channels (right, shown in a two-step magnifying glass style to bridge the differences in length scales). The active components can be amphiphilic molecules or complexes residing in the bilayer membranes forming between adjacent droplets, and  carefully chosen contents of the droplets themselves.\label{Concept}}
\end{figure}

The concept which thus emerges, and on which we will dwell in the present article, is as follows. We create a system of microfluidic channels, compartments, posts, orifices and so on, on length scales which are well manageable by conventional top-down lithography. This system is then filled with soft molecular materials, which self-assemble within the prescribed geometry, forming well-defined structures down to much smaller scales. To be more explicit, we envisage the use of mono-disperse water-in-oil emulsions, with a small volume fraction of the oil phase and a suitable surfactant (e.g., a lipid) for stabilization. This type of emulsions, where the continuous phase is very dilute, are commonly called gel-emulsions \cite{Solans1998}. In an externally predefined channel geometry, the water droplets will form well-oriented crystalline arrangements \cite{Garstecki2004,Seo2005}. Under suitable conditions, the interfaces of every two adjacent droplets will form molecular surfactant bilayers. These can serve, by means of dispersion forces and wetting, as nanoscopic tweezers for holding active components, such as ion channels or smaller molecules with non-trivial electronic properties. For molecular electronic circuitry, the water droplets take the role of the 'solder points'.

It is thus inherent to the concept outlined above that we are using self-assembly processes on two completely different scales: the scale of the droplets within the prescribed geometry of the micro-fluidic channel structures, and the nanoscopic scale of the membrane thickness and the functional molecular building blocks. On both scales, self-assembly is then purely driven by wetting forces. As will be discussed below, these forces can not only be used for maintaining certain membrane structures, but also for their controlled manipulation.

Clearly, there will be any gain over conventional 'top-down' devices only if the self-assembled soft matter structures attain length scales below $50 {\rm nm}$ or so. In the present study, we still keep far away from this scale, in order to facilitate the production and observation of the structures and processes at this very preliminary stage. The present article rather aims at identifying basic mechanisms and conceptional elements which appear promising, but are still to be scaled down considerably. It should be noted, however, that one of the major potential problems of the concept, the instability of emulsions against coalescence, has been found to fade away as size is reduced. Some results pointing in this direction will be briefly discussed. 

\section{formation and stability of membranes in micro-fluidic systems}

It has been recognized long ago that for micro-fluidics to unfold
its full power, it is advantageous to use two immiscible liquids as
the working medium instead of a single liquid phase
\cite{Whitesides2006}. Not only can isolated droplets immersed in a
continuous liquid be used as compartments for controlled reactions
\cite{Ismagilov2006}, but the complex interplay of the liquid/liquid
interface with the channel geometry and hydrodynamic pressure
gradients gives rise to an abundance of effects which can be
harnessed into the engineering of functional devices. So far, this
has been mainly limited to the use of isolated aqueous droplets in
an oily continuous phase, as containers for chemical, biochemical,
or biological processes. In these devices, the droplets are
transported within the channel mainly due to their hydrodynamic drag
in the surrounding oil phase. Contact between adjacent droplets was
not desired, in an effort to keep droplet coalescence and other
cross-talk effects to a minimum.

The situation becomes completely different if the volume fraction of
the oil phase is reduced such that droplets maintain mutual contact
throughout. The transport of the droplets in the device is then not
anymore determined by the streamlines of the oil phase, but by the
(dynamically varying) geometry of optimum packing of the droplets
within the channel geometry \cite{Wiebke,Priest2006}. In fact, the
arrangement of spherical droplets has been reported to change from
random to crystalline as their volume fraction becomes large. In a
square channel with a lateral dimension of about 4 droplet
diameters, a volume fraction of 0.75 shows a crystalline order,
while a random arrangement is observed at 0.55 \cite{Priest2006}.

It is clear that this opens up qualitatively new possibilities of
droplet manipulation. In particular, the existence of several
meta-stable configurations gives rise to strong hysteresis effects
in the droplet geometry. Droplet motion may thus not be reversible,
and become strongly history dependent. If the droplets are all of
the same size, i.e., if a mono-disperse gel emulsion is used, these
effects can in principle be exploited to dynamically access a large
variety of droplet configurations in the
device\cite{Garstecki2004,Seo2005,Fuerstman2007,TuulAPL09}.
Furthermore, these configurations are geometrically rather stable,
because the wetting forces determining the angles of contact of the
oil lamellae spanning between the droplets, give rise to substantial
energy barriers for any spontaneous rearrangement.

\begin{figure}[h]
\includegraphics[width = 8.5cm]{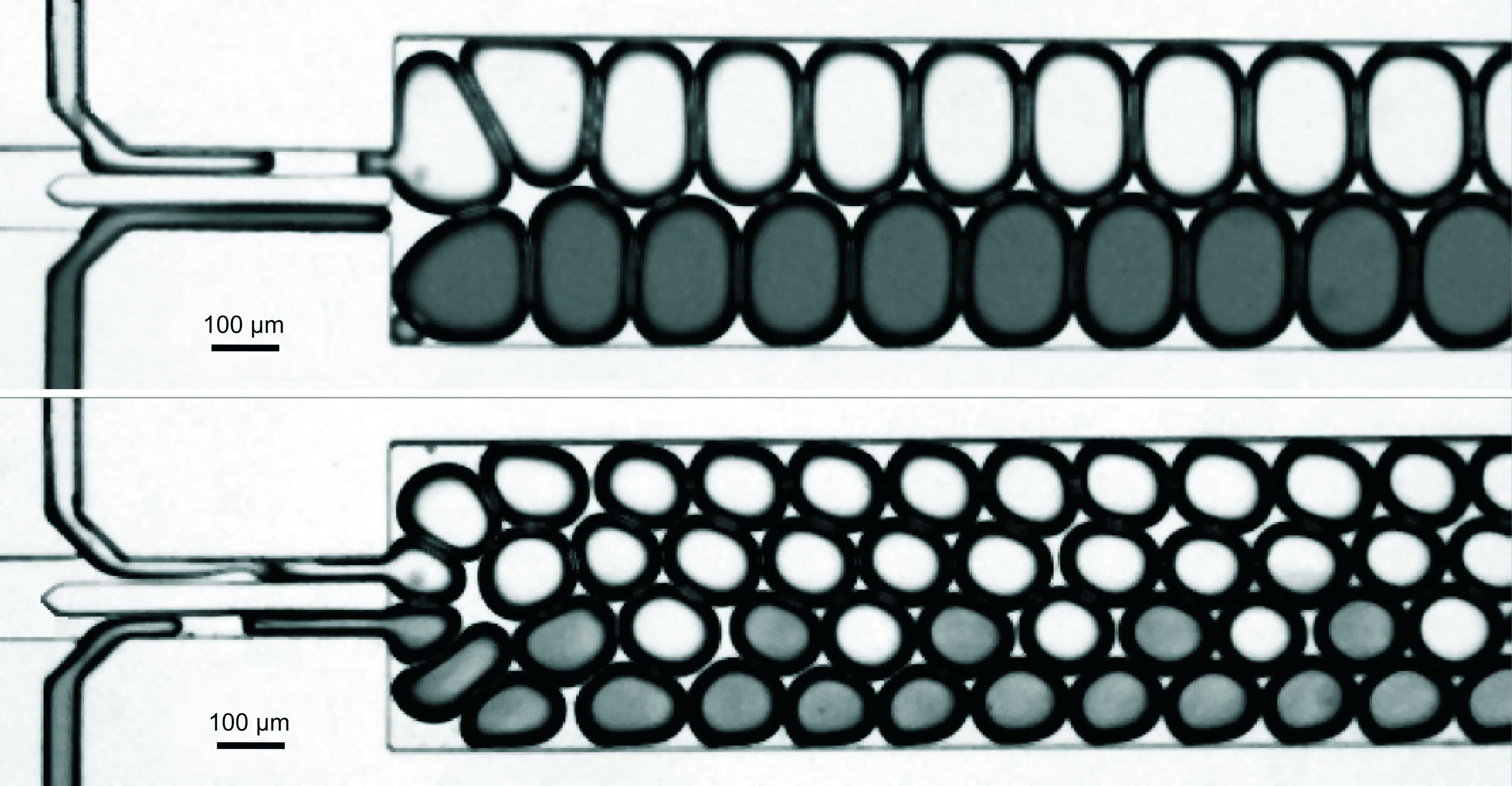}
\caption{Gel emulsions of water in oil generated by step emulsification, forming rafts in a wider channel. The channel width is 500 $\mu$m, the volume fraction of the aqueous phase is 0.75. The elementary cells of such rafts can be quite complex, as the lower example shows. Controlled rearrangement of these configurations are possible by appropriately chosen channel geometries \cite{Priest2006,Tuul2009,TuulAPL09}. \label{ElementaryCell} }
\end{figure}

A simple example is displayed in Fig.~\ref{ElementaryCell}, which
shows regular configurations of droplets with different content. The
channel structure was formed in poly-dimethyl-siloxane (PDMS, Dow
Chemicals) by standard soft lithographic techniques. We used
squalane with 15 mg/ml of
1,2-diphytanoyl-sn-glycero-3-phosphocholine (DPhyPC, Avanti Polar
Lipids) plus 15 mg/ml of cholesterol as the oil phase, which enters
the emulsification unit \cite{Venkat2008} from the left. For
coloring the aqueous phase injected in the lower side channel, we
used 2 mM DPhyPC doped with 2 molar \% of
1,2-dioleoyl-sn-glycero-3-phosphoethanolamine-N-(carboxy
fluorescein) in the upper picture, and
1,2-dipalmitoyl-sn-glycero-3-phosphoethanolamine-N-(lissamine
rhodamine B sulfonyl) (both dyes from Avanti Polar Lipids) in the
lower picture. The dye lipid was sonicated in the aqueous phase to
create liposomes. We see that the droplets form a periodic structure
with an 'elementary cell' withstanding transport along the channel.
The relative positions of the droplets within the arrangement are
fixed, as is obvious due to the droplet color.

Most of the results discussed here were obtained with channel
dimensions on the order of 100 microns or more, for the sake of
simplicity of generating the devices and optical transparency.
However, we repeated some of the experiments with channels etched in
silicon which were much narrower (down to 20$\mu m$ wide). We found
that the smaller the droplets and channel dimensions were, the more
stable were the membranes. A straightforward explanation is that the
Laplace pressure, which scales as the inverse of the droplet radius,
sets the energy scale for any disturbance leading to substantial
rearrangements of deformations of membranes, and thus potentially to
coalescence. Extrapolating this general trend, further down-scaling
is expected to yield emulsion structures which are even much more
stable. It should be noted that the formation of stable,
mono-disperse emulsions with droplet diameters of 50 nm or less is
routinely possible by suitable methods \cite{Land2009}.

%

\begin{figure}[h]
\includegraphics[width = 8.5cm]{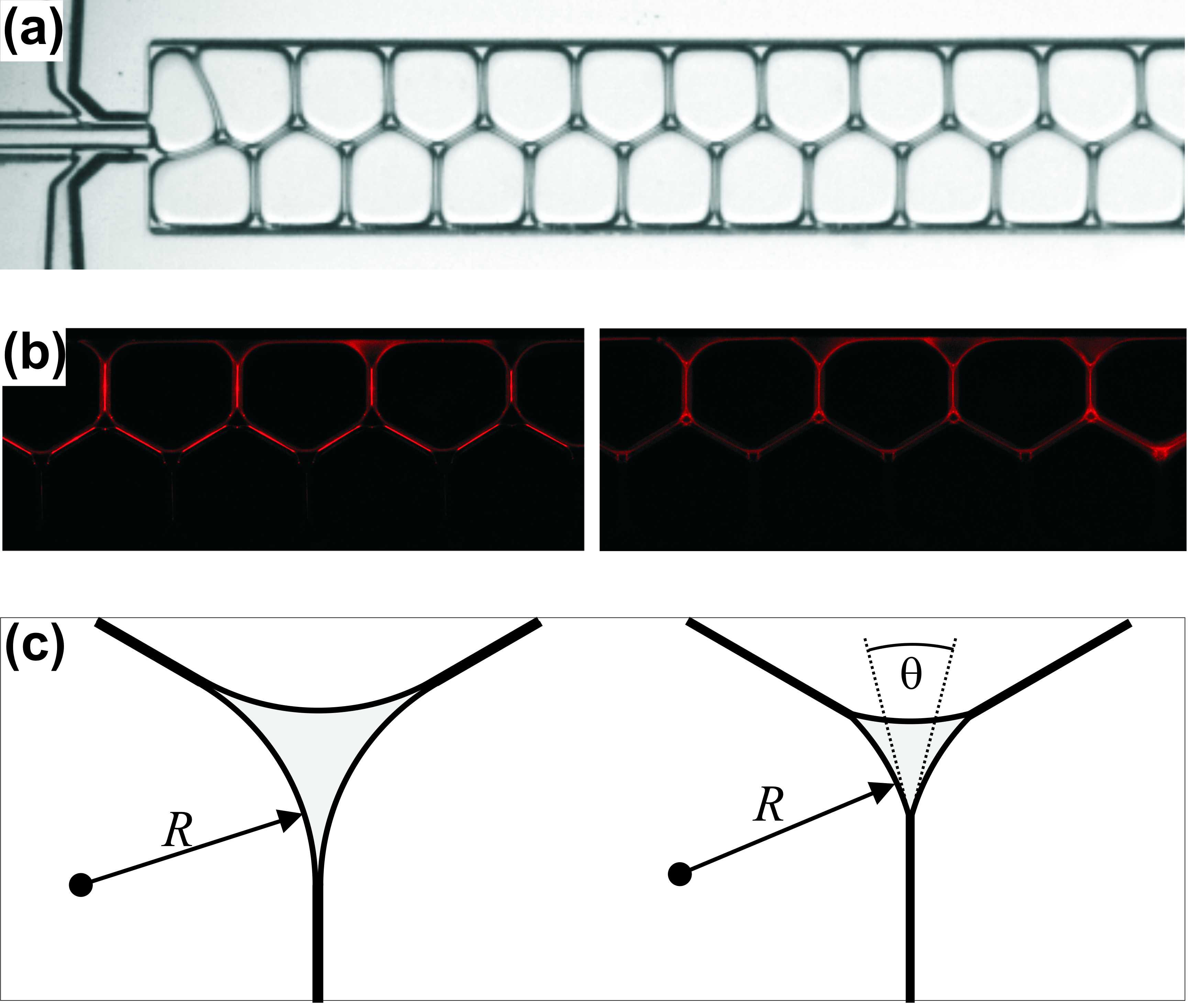}
\caption{(a) Gel emulsion in a channel (500 $\mu$m wide) at very low volume fraction of the continuous oil phase (below 10\%). (b) Micrograph of an emulsion like the one shown in (a), taken at the fluorescence wavelength of the dye (di-4-anepps) which was fed into the upper row of droplets. Illumination is by UV light. Left: as prepared. Right: a few seconds later, when membranes have formed. (c) Geometry of Plateau borders, where three oil lamellae (or membranes) meet at 120 degrees. Left: as prepared. The droplet surfaces meet tangentially. Right: after membranes have formed. The contact angle, $\theta$, at the end of each membrane has attained a finite value. \label{ZigzagMembranes} }
\end{figure}

Spontaneous membrane formation is well known to occur if two
surfactant-laden oil/water interfaces are brought in intimate
contact on their lipophilic sides \cite{Anders1983}. Recently, this
has also been demonstrated with aqueous drops placed on a substrate
next to each other in an oil background phase
\cite{Holden2007,Heron2007}. It is to be expected that a similar
process will occur if a suitable gel emulsion is squished into a
micro-fluidic channel at sufficiently low continuous phase volume
fraction. That this is indeed the case is demonstrated in
Fig.~\ref{ZigzagMembranes}. Here we have produced a zig-zag
structure of mono-disperse aqueous droplets by step emulsification
\cite{Venkat2008}. The oil is squalane, with the lipid mono-olein as
the surfactant at a concentration of 25 mM/l, which is well above
the critical micelle concentration (CMC). The droplets in the upper
row contained a fluorescent dye (di-4-ANEPPS, Invitrogen) which
preferentially enters the central lipophilic zone of a lipid
bilayer. In Fig.~\ref{ZigzagMembranes}a, which was taken immediately
after the formation of the droplets, bright lines of fluorescence
are visible in the oil layers extending between the droplets of the
upper row. Fig.~\ref{ZigzagMembranes}b is taken a few seconds later.
Clearly, some of the bright lines have disappeared, indicating the
expulsion of the majority of the oil from between the droplets. The
transition from the bright line to this faint glow occurs abruptly,
and for each oil layer independently. It suggests itself to
interpret this transition as the formation of a lipid bilayer
separating adjacent aqueous droplets.

\begin{figure}[h]
\includegraphics[width = 8.5cm]{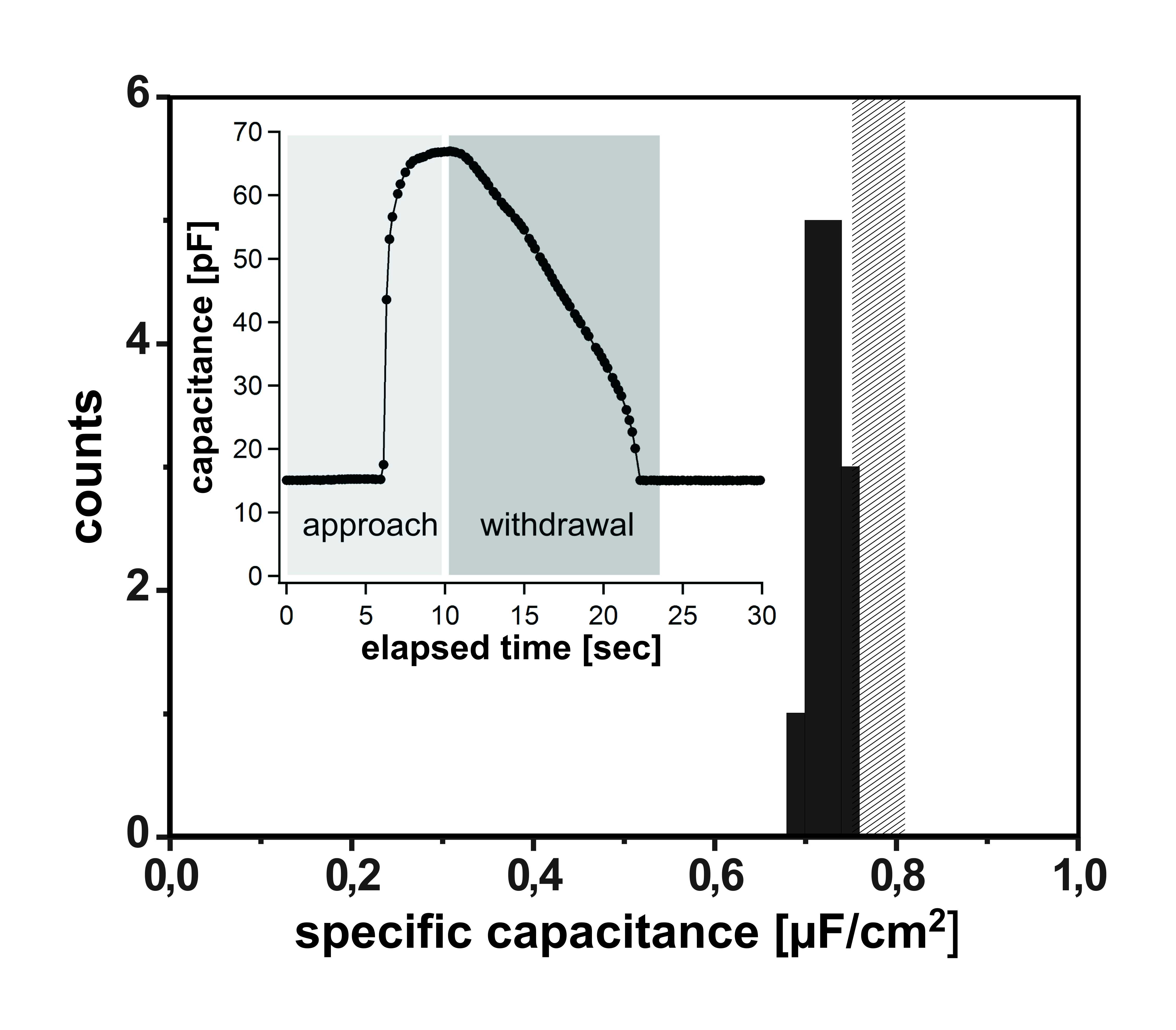}
\caption{Measurements of the specific capacitance of a bilayer membrane of mono-olein in squalane. The black histogram represents the measured values. The grey bar indicates the literature values for oil-free mono-olein membranes. Inset: trace of capacitance measurement upon approach and withdrawal of two droplet surface to/from each other. \label{CapacitanceJump} }
\end{figure}

That this is indeed the case is shown in Fig.~\ref{CapacitanceJump}.
For this experiment, two droplets were used which contained 150 mM/l
NaCl in Millipore water, a content similar to those in
Fig.~\ref{ZigzagMembranes} except for the dye. These were gradually
approached, in an oil phase consisting of 25 mM/l mono-olein in
squalane, beyond the formation of a contact between their
interfaces, such that the latter formed a flat region separating the
droplets. After a few seconds, the same abrupt transition was
observed as reported in Fig.~\ref{ZigzagMembranes}. This time the
droplets were connected via electrodes to a patch-clamp amplifier,
such that the capacitance could be continuously measured. The inset
in Fig.~\ref{CapacitanceJump} shows a trace of the sample
capacitance for contact formation and subsequent withdrawal. From
microscopic inspection of the flattened region of the interface, one
can estimate the diameter of that region, and thereby its area. This
allows to calculate the specific capacitance of the membrane thus
formed. The histogram displayed in the main panel summarizes a
series of experiments performed with the same pair of droplets.
Clearly, the specific capacitance is well reproducible. The hatched
region indicates estimates for a solvent-free membrane of
mono-olein, and compares favorably with our results. The slight
deviation may be either due to systematic errors in the estimation
of the area, or to some residual oil trapped in the membrane formed.

The lines at which three membranes meet, at an angle of 120 degrees
at equilibrium, is called a Plateau border \cite{Wiebke}. These
objects are clearly visible in Figs.~\ref{ZigzagMembranes}a
and~\ref{ZigzagMembranes}b. In the central image, they show up as
dark regions surrounded by the lines of glow from the surfactant
layers and the oil. It is clear that as long as there is no
membrane, the surfactant layers delimiting adjacent droplets will
meet tangentially, i.e., at zero contact angle. This is illustrated
in Fig.~\ref{ZigzagMembranes}c, where the grey shaded area
corresponds to the dark Plateau border in the fluorescence
micrograph. The latter reveals that upon membrane formation, the
size of this region shrinks significantly. We can use this effect to
determine the contact angle at which the surfactant layers meet the
membranes, as illustrated on the right hand side of
F.g~\ref{ZigzagMembranes}. For the shaded area, $A$, one readily
finds
\begin{equation}
A = R^2 \left(\sqrt{3}\cos\frac{\theta}{2} + \frac{3}{2}(\theta-\sin\theta)-\frac{\pi}{2}\right)
\end{equation}
The radius of curvature, $R$, is directly linked to the pressure,
$p$, in the droplet via $p = \gamma/R$, where $\gamma$ is the free
energy per unit area of the surfactant-laden water/oil interface.
$R$ can thus be considered to remain unchanged as the membrane
forms. It is then easy to obtain the contact angle, $ \theta$, from
the size of the Plateau borders. We obtain $\theta = 48\pm 3$
degrees.

At the three-phase contact line, where the membrane meets the adjacent water/oil interfaces, force balance yields
\begin{equation}
\Gamma = 2\gamma \cos\frac{\theta}{2}
\end{equation}
where $\Gamma$ is the free energy per unit area of the membrane. We
can thus directly infer the membrane free energy from inspection of
the Plateau borders if $\gamma$ is known. Using the standard pendant
drop method, we obtained $\gamma = 1.77 \pm 0.11$~mN/m. Our result
for the membrane free energy is thus $\Gamma = 3.23  \pm 0.20$~mN/m.
The formation of the membrane is therefore accompanied by a gain in
free energy of $\Delta F = 2\gamma - \Gamma = 0.31 \pm 0.02$~mN/m.

The interfacial free energies involved are of interest for the
potential performance of devices designed in the way proposed here.
In order to take full advantage of the concept, droplets have to be
formed and moved relative to the geometry provided by the
micro-fluidic channel structure. The excess free energy of the
droplet interfaces sets the scale for the pressures to be applied in
these processes. The Laplace pressure is given by $p_L \approx
\gamma/l$, where $l$ is the size of the smallest orifice to be
passed. For $\gamma = 3$~mN/m, we obtain $p_L = 1$ bar for a 30 nm
orifice, which appears well feasible.

\begin{figure}[h]
\includegraphics[width = 8.5cm]{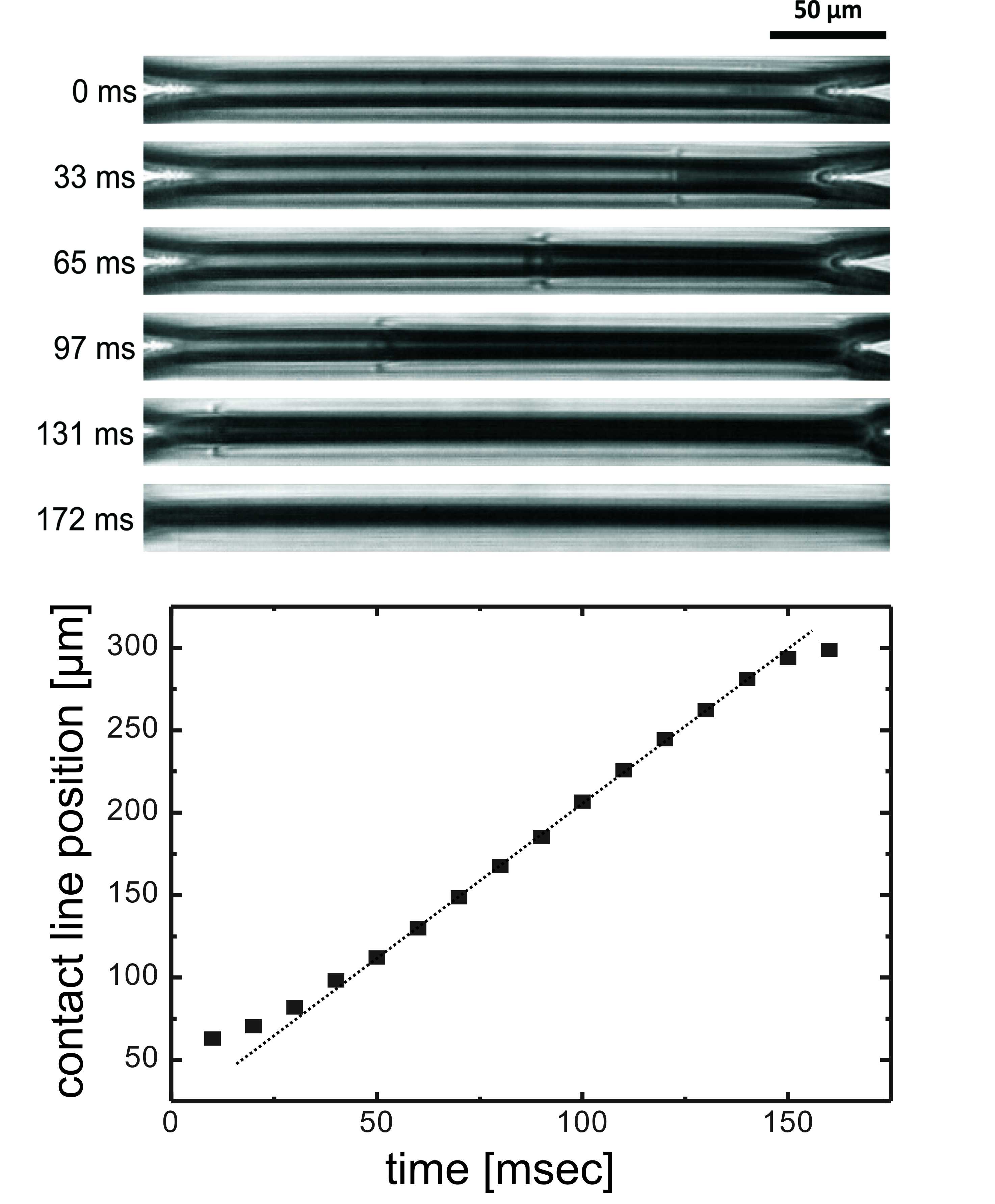}
\caption{Time-resolved micrograph of membrane formation in a micro-fluidic system. Top: time-lapse images of three-phase contact line moving across the oil lamellae while membrane is forming. The total diameter of the membrane is 300 $\mu$m. Bottom: Three-phase contact line position as a function of time. The slope of the dotted line is 1.9 mm/s. \label{ZipperDynamics} }
\end{figure}

As the sharp rise in the capacitance trace shown in the inset of
Fig.~\ref{CapacitanceJump} suggests, the formation of the membrane
is a rapid process. Fig.~\ref{ZipperDynamics}a displays a series of
images captured with a high speed camera (1000 frames per second,
Photron SA 3). The zipper-like transition is clearly discernible,
although details on the scale of the membrane thickness are of
course not accessible to optical imaging. The total time of
formation in this run was about 150 ms. Fig.~\ref{ZipperDynamics}b
shows the membrane diameter as a function of time, as observed
during a single formation event. The approximate constancy of the
contact line velocity (i.e., the first derivative of the diameter
with respect to time) suggests a constant power of dissipation,
which is due to the viscous friction in the vicinity of the edge of
the membrane, where the two aqueous phases and the oil phase meet.
More specifically, we can compare the contact line velocity of about
1.9 mm/s with the capillary velocity, $v_c = \gamma/\eta =$ 4.1 cm/s
, where $\eta$ is the viscosity of the liquid (in this case Squalane
which has a viscosity of 43.4 mPa.s). This is a natural velocity
scale for the system, corresponding to the balance of interfacial
and viscous forces. Clearly, the measured contact line velocity is
much less than $v_c$, which can be attributed to the diverging
viscous stress near the three-phase contact line \cite{Dussan1979}.

%

For an effective manipulation of the droplet configuration within the micro-fluidic setup, it may be necessary to move the emulsion through the micro-fluidic device. This can be done either externally by  pressure or volume control (e.g., syringe pumps), or internally by means of suitable local mechanisms. In any case, it is of central importance that these manipulation steps do not lead to the destruction of membranes. One might anticipate that this constraint poses a serious conceptual problem, since the membranes are objects of minute thickness (about 4.5 nm in the case of mono-olein) and relatively small energy of formation (see above).

\begin{figure}[h]
\includegraphics[width = 8.5cm]{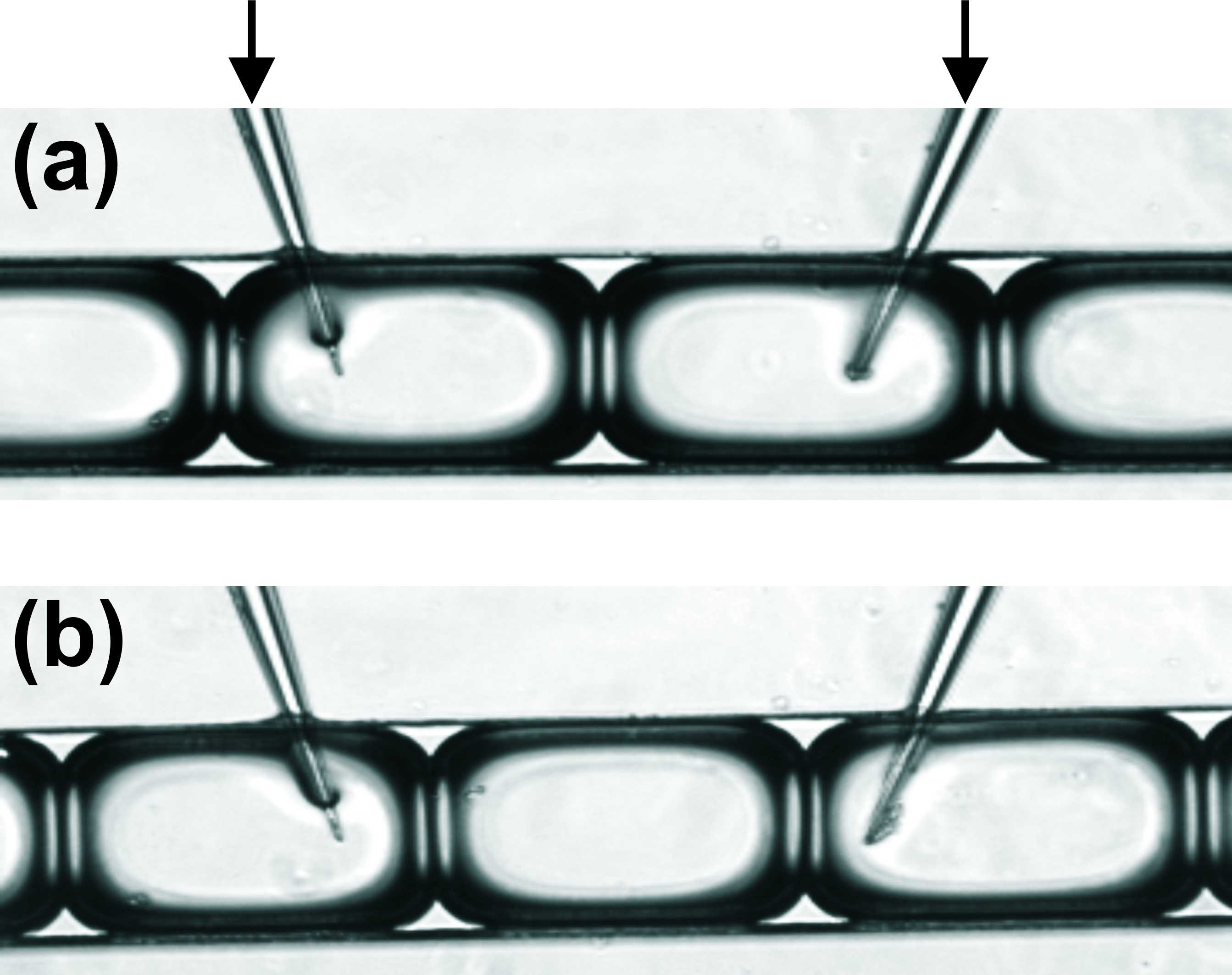}
\caption{Contacting droplets (150 mM/l NaCl in Millipore water) electrically by means of micro-pipettes poked through the walls of the PDMS-micro-device. The pipettes are equipped as ${\rm Ag/AgCl_2}$ electrodes and connected to a patch-clamp amplifier. The oil phase consists of squalane plus 25 mM/l mono-olein. The channel is 100 $\mu$m wide. \label{MicroPipettes} }
\end{figure}

Quite remarkably, we found the bilayer membranes to be very stable against mechanical stresses, such as those exerted on them in micro-fluidic flow. A particularly striking example is shown in Fig.~\ref{MicroPipettes}. Glass micro-pipettes can be inserted through thin walls (100 $\mu$m) created in the PDMS directly into the droplets. This is important for establishing ohmic contacts to the aqueous droplet content by means of standard ${\rm Ag/AgCl_2}$ electrodes, as used in electro-physiology. In order to isolate the droplets hydrodynamically from the inside of the pipettes, the latter were filled with Millipore water plus 150 mM/l of NaCl and a small amount of agarose to form a gel. Electrical contact was established by inserting a 500 $\mu$m diameter Ag wire into the pipette, which was chlorided before electrochemically with 3 M KCl. The droplets were pumped through the channel at a rate of about three droplets per second. As seen from Fig.~\ref{MicroPipettes}a and Fig.~\ref{MicroPipettes}b, the membranes survive their motion past the inserted glass pipettes: coalescence is not induced. We found that the stability of the membranes against such mechanical stress depended on the type of intruder. While the glass pipette electrodes did not seem to affect the membranes in any way, a tungsten wire (100 $\mu$m diameter, sharpened to a 30 $\mu$m tip) induced immediate coalescence.

\section{electrical contacts and characteristics}

\begin{figure}[h]
\includegraphics[width = 8.5cm]{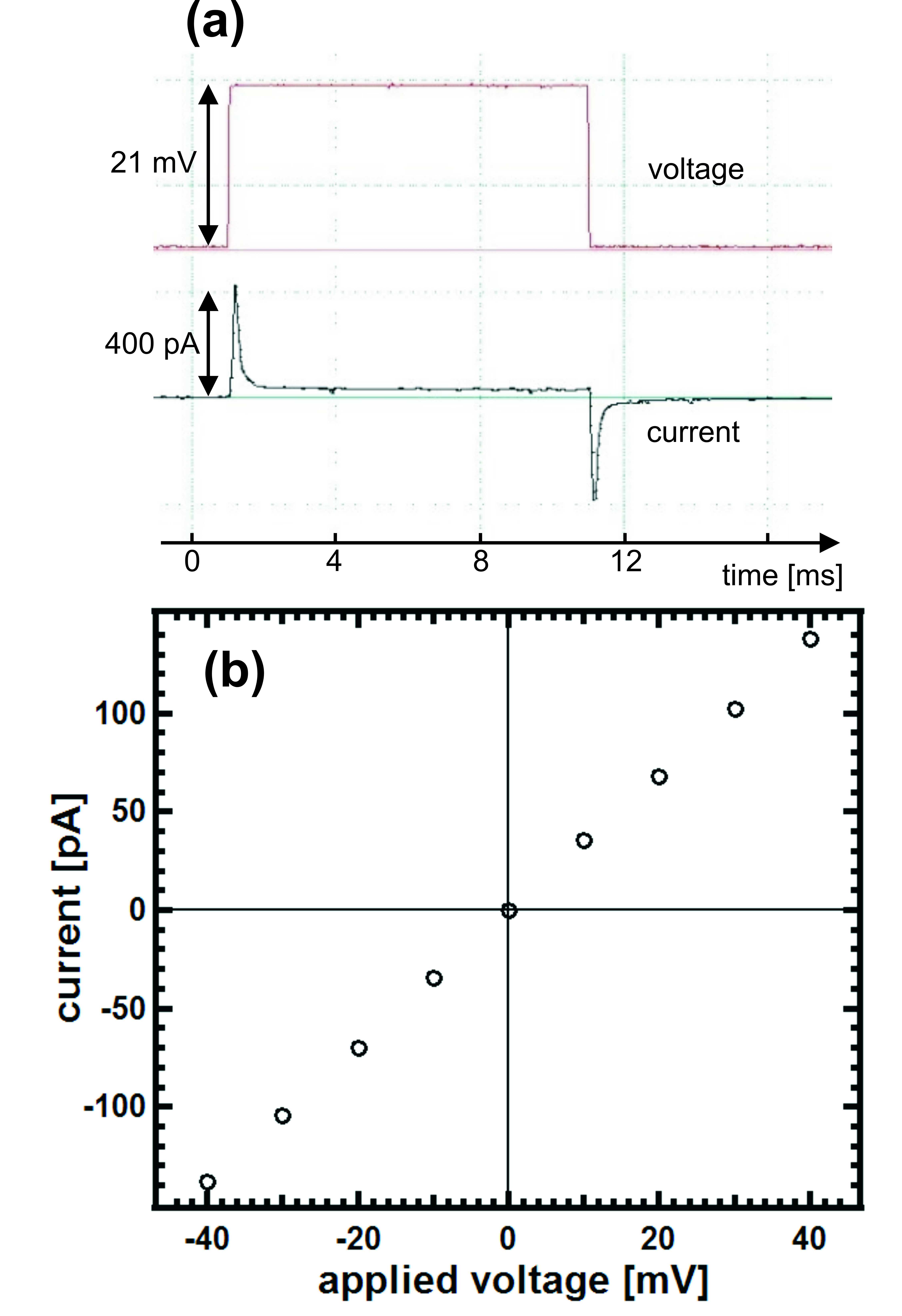}
\caption{Electrical properties of mono-olein membranes formed and suspended in a micro-fluidic channel. (a) A square wave voltage is applied (top). The current trace (bottom) shows distinct peaks indicating the charging current of the membrane capacitance. (b) The small offset observed in the current trace in (a) is plotted as a function of applied voltage. We clearly see an ohmic behavior, which probably stems from ionic impurities. \label{VoltageCurrentTrace} }
\end{figure}

Let us now turn to the possibility of constructing electrical circuitry out of gel emulsions in micro-fluidic systems. The first thing we have to demonstrate is that the aqueous droplets, which are now to be considered the 'solder points' of potential self-assembled circuits, can be suitably connected to external leads. The setup shown in Fig.~\ref{MicroPipettes} can indeed be used to measure the electrical properties of the membranes spanned between adjacent droplets. Fig.~\ref{VoltageCurrentTrace}a shows typical traces obtained upon applying a square wave to a single membrane through the pipettes, which were configured as ${\rm Ag/AgCl_2}$ electrodes. Both the current and the voltage are recorded using a standard patch clamp amplifier (HEKA, EPC 10).  We clearly see the loading current of the membrane capacitance, as well as an ohmic current persisting as long as the voltage is applied.

While the capacitive signal is unambiguously due to the membrane capacitance and corresponds to the specific capacitance of about $0.7 {\rm \mu F/cm^2}$ (c.f. Fig.~\ref{CapacitanceJump}), the ohmic resistance is masked by the resistance of the ionic conductance of the surrounding aqueous solution. Comparing the traces obtained with the configurations displayed if Figs.~\ref{MicroPipettes}a and b, we obtain the 'pure' ohmic characteristic of a single membrane, as displayed in Fig.~\ref{VoltageCurrentTrace}b. This conduction is probably due to ionic impurities which are sufficiently lipophilic to cross the membrane \cite{Dilger1985} in a thermally activated process.

\begin{figure}[h]
\includegraphics[width = 8.5cm]{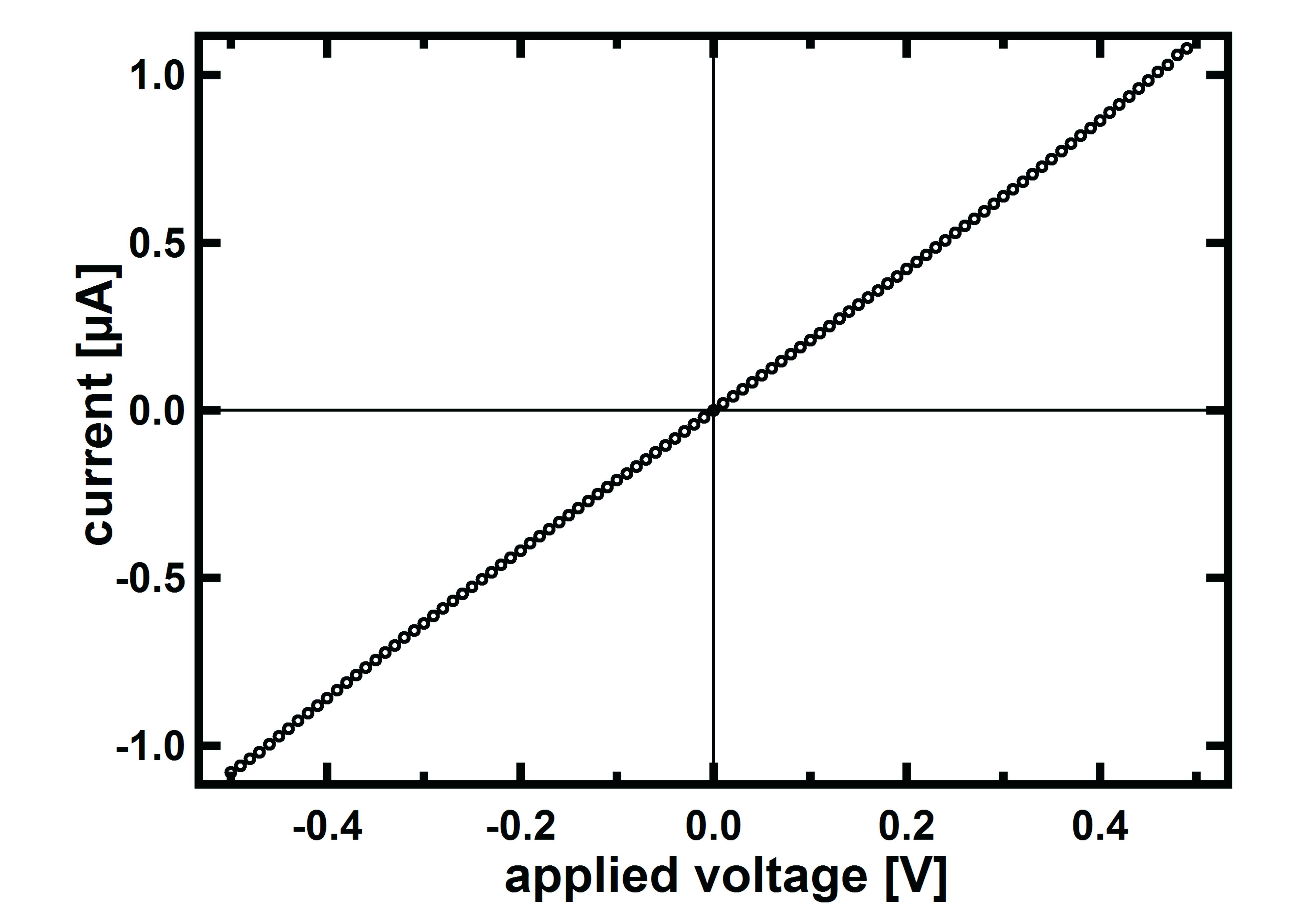}
\caption{Current/voltage characteristic of a mono-olein membrane doped with Gramicidin A ion pores.\label{Gramicidin} }
\end{figure}

Next we want to demonstrate that the capability of membranes to incorporate active components, as it is well known from experiments in membrane physiology \cite{Hwang2007,Holden2007}, also pertains to the micro-fluidic setting. We have therefore added gramicidin ion pores to the liquid phase in the device, in order to see whether the presence of the channel walls might hamper their performance noticeably. Fig.~\ref{Gramicidin} shows the current-voltage relationship of a membrane doped with gramicidin A ion pores (Sigma-Aldrich).  The gramicidin is added to the aqueous phase at a concentration of $1.1 {\rm \mu M}$. Clearly, there is now a dramatically increased conductance of the membrane, which we attribute to the ionic conductivity of the pores.  From the unit conductance of the gramicidin A ion pore \cite{Hladky1972}, we estimate the number of pores in our membrane to be in the order of $10^6$. Since the membrane is now significantly conducting as compared with the native state, voltages of 500 mV (and even up to 1 V) can easily be applied without rupturing the membrane. In the native state, we found that applied voltages beyond roughly 300 mV ruptured the membrane.

\begin{figure}[h]
\includegraphics[width = 8.5cm]{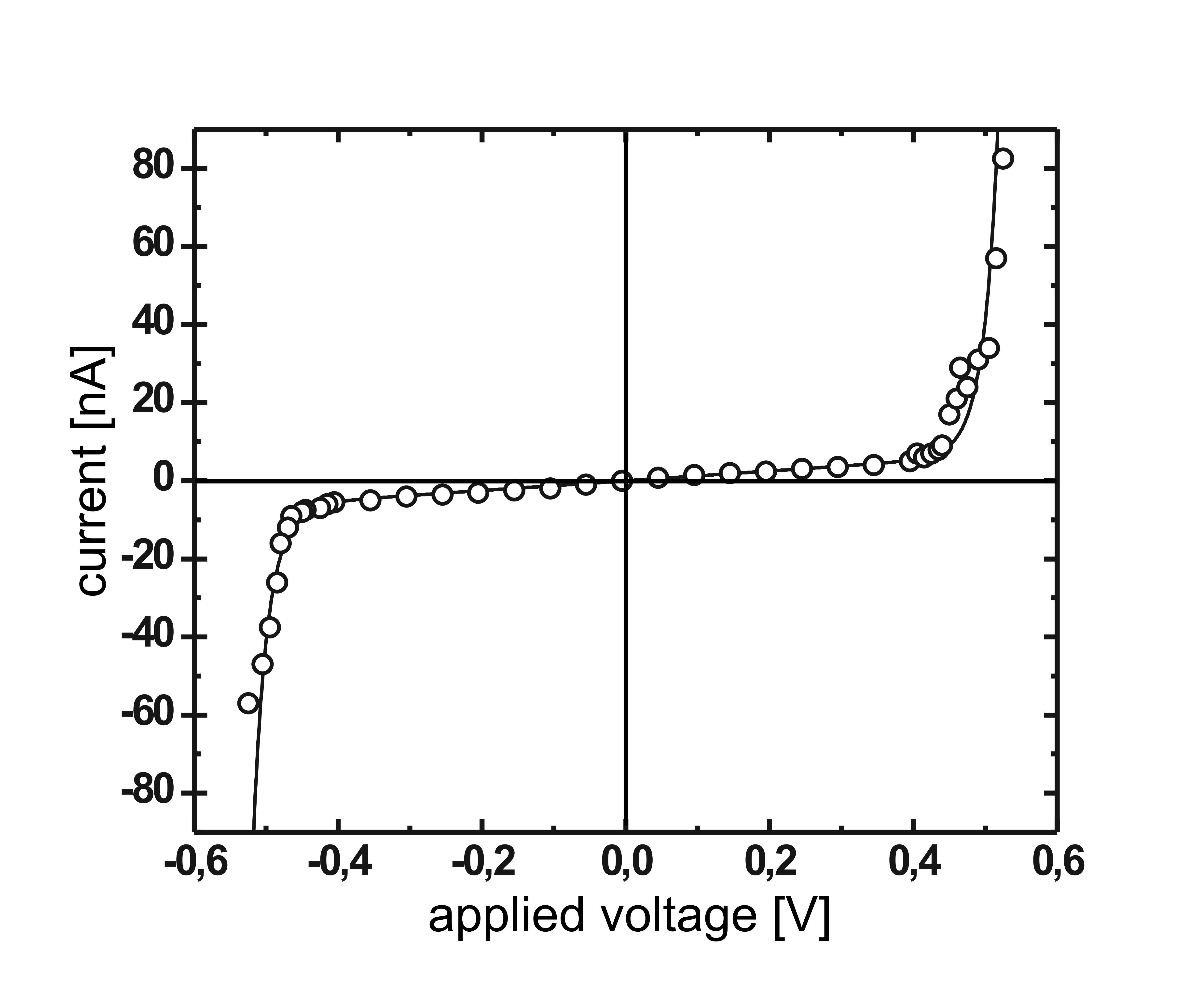}
\caption{Current/voltage characteristic of a mono-olein membrane. Aside from an ohmic central part, there is a pronounced rise in the current at a threshold voltage of about 500 mV. This is attributed to reversible electroporation. The solid curve corresponds to what would be expected theoretically in that case.  \label{Electroporation} }
\end{figure}

Along with the applied voltage, the width of the applied voltage pulse is also crucial in studying the current/voltage characteristic of a membrane. The result of such an experiment on a native membrane similar to the one described above is displayed in Fig.~\ref{Electroporation}. In comparion with the previous experiment (pulse width of 10 ms), here the applied pulse is much shorter and is of 1 ms duration. We observe a clearly defined threshold voltage above which the current rises sharply. We interpret this as reversible electroporation due to electric-field induced formation of nano-scale liquid necks within the membrane \cite{Weaver2003,Weaver2006}. Also, it is known from studies on cells that pore formation is favoured when the electric pulses are relative short and typically below 1 ms \cite{Xie1992}. The rate of creation of (transient) nano-pores due to the applied electric field is believed to scale as $\exp(U^2/kT)$, where $U$ is the applied voltage \cite{Weaver2006}. Accordingly, the solid curve has the form
\begin{equation}
I = a U + I_0 \sinh\left(\frac{b U^3}{\mid U \mid}\right)
\end{equation}
where $a$ accounts for impurity-induced ohmic conductance. For the constant $b$, we obtain ${\rm 49 kT/V^2}$, which is a reasonable value for bilayer membranes \cite{Weaver2006}. The sharp rise of the current at the critical voltage of about 500 mV shows that the membrane can be used as a voltage stabilization device, similar to a Zener diode, without further manipulation.

\section{coupling chemical oscillators via membranes}

So far, we have considered the aqueous droplets merely as electrically conducting links between the membranes, which so far were the main players in the concept. Now we want to turn to more complex contents of the droplets themselves. More specifically, we will consider droplets acting as chemical oscillators, and investigate how they may be coupled to each other by the bilayer membranes separating them.

The most thoroughly studied oscillating chemical reaction is the so-called Belousov-Zhabotinski (BZ) reaction \cite{Zhabotinsky1964}. There are several formulations which have been used. Most of them contain an autocatalytic step decomposing bromate (${\rm BrO_3^-}$) as the main educt, with the help of an oxidizer, such as ${\rm Fe^{III}}$ ions. The reduction of the ${\rm Fe^{III}}$ to ${\rm Fe^{II}}$ provides an optical indicator for this step, which can be enhanced by using the ferroin complex as a carrier for the Fe ions. The progress of the reaction is then accompanied by a color change from deep red to light blue. The product, ${\rm BrO_2^-}$, acts as a catalyst, or 'promotor' in this first step. A second step involving malonic acid is coupled to this system, which by re-oxidizing the ${\rm Fe^{II}}$ to ${\rm Fe^{III}}$ produces bromine ions. The latter react with the ${\rm BrO_2^-}$ promotor to yield ${\rm BrO^-}$ as the final product. The bromine thus scavenges the promotor of the autocatalytic step, and is therefore considered as the 'inhibitor'.

After having been almost unrecognized after its discovery for many years, the BZ reaction has later become one of the paradigm systems for dynamical systems and self-organization. In spatially extended settings, the BZ reaction gives rise to complex propagating wave patterns, which are reminiscent of spatio-temporal patterns known from other catalytic systems driven far off thermal equilibrium \cite{Ertl1994,Ertl1993}. In small containers, however, where diffusive coupling across the container is strong, they behave like single homogeneous oscillators. It then suggests itself to consider larger ensembles of coupled BZ oscillators as analogous to, e.g., systems of firing neurons.

It has been shown before that systems like that can be nicely realized by using aqueous droplets containing the BZ educts as the oscillators, and coupling them via an oil phase separating the droplets \cite{Toiya2008}. Both the promotor and the inhibitor are sufficiently hydrophobic to enter the oil phase easily, such that they can diffuse from one droplet to the other if they are sufficiently close. The delicate balance of promotive and inhibitory coupling gives rise to a wealth of oscillation patterns \cite{Toiya2010}. They strongly differ from the wave-like patterns which are well-known from continuous systems, and are as yet not fully understood. For the concept we are considering here, it is of great interest whether we can distinguish coupling through the bulk oil phase from coupling through a bilayer membrane, and to investigate what types of coupling can be achieved solely via the latter.

We prepared emulsions consisting of droplets containing the BZ educts in aqueous solution, suspended in an oil phase consisting of squalane, with mono-olein at concentrations well above the critical micelle concentration (CMC). The mono-olein serves two purposes. First,it forms dense surfactant layers at the oil/water interface and readily form bilayer membranes, as shown above. Second, the C=C double bond in the mono-olein molecule acts as an efficient scavenger for bromine, since the latter rapidly reacts with this site. The oil phase is thus expected to efficiently suppress coupling between neighboring droplets. As soon as a bilayer has formed, however, coupling is expected to commence rapidly. Most of the mono-olein molecules forming the membrane have been exposed to the reaction products of the BZ system before and will thus be brominated already when the membrane forms. But even if the latter was formed with fresh mono-olein, it would be readily saturated with bromine given its minute thickness, and thus would let further bromine compounds pass easily.

\begin{figure}[h]
\includegraphics[width = 8.5cm]{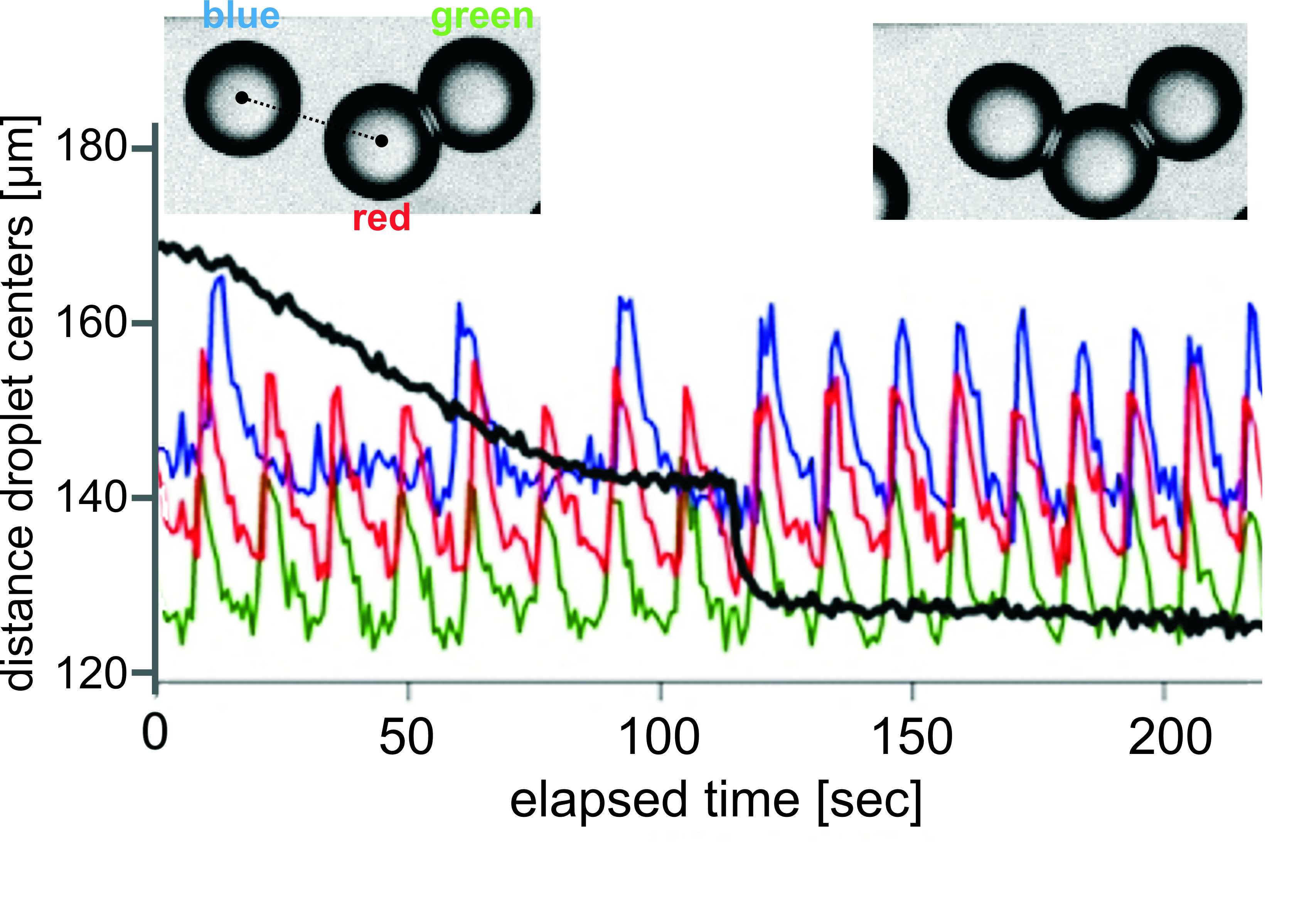}
\caption{Effect of membrane formation upon the phase coupling of chemical oscillators. The blue, red, and green trace represent the transmittance of the three droplets shown in the insets as a function of time. The black curve shows, on the same time axis, the distance of the 'blue' from the 'red' droplet (measured center-to-center). Clearly, the oscillations couple in phase as soon as the membrane is formed (jump in the black curve), but not before. \label{MembraneCoupling} }
\end{figure}

In order to demonstrate membrane-specific coupling, we have observed
the oscillation behavior of BZ droplets while they were diffusively
moving relative to each other, finally forming bilayer membranes.
Fig.~\ref{MembraneCoupling} shows the transmittance traces of three
droplets, exhibiting the spike pattern characteristic of the BZ
oscillation. The auto-catalytic reaction step which reduces the
strongly absorbing ${\rm Fe^{III}}$ to the almost clear ${\rm
Fe^{II}}$ solute leads to a steep increase of the transmittance.
This is followed by a smooth decrease due to the gradual
re-oxidation of the ${\rm Fe^{II}}$ involving the malonic acid.
Initially, only the two droplets whose transmittance traces are
shown in red and green are connected by a bilayer membrane. This was
known from the direct observation of the membrane forming process.
The third droplet, the transmittance trace of which is shown in
blue, had some distance to the first pair, with about 50 microns
surface separation from the 'red' droplet. A large oil volume
fraction was used in this sample, such that the droplets could
diffuse freely for some distance. Clearly, the red and green
transmittance traces are phase locked, while the blue trace follows
its own pace, showing no sign of influence from the other two for
the first $\approx 100$ seconds shown. The black curve represents
the distance of the centers of the 'red' and 'blue' droplets, as
determined from fitting circles to their images. It shows how the
'blue' droplet gradually drifts towards the 'red'. At around 100
seconds, the surfaces of the droplets have come so close that the
drift is stopped due to the diverging hydrodynamic resistance of the
flat sphere-to-sphere contact. At about 113 seconds, the droplet
centers are rapidly pulled together, which we interpret as the
formation of the bilayer membrane (cf. Fig.~\ref{ZigzagMembranes}).

In order to verify this interpretation, we can estimate the contact
angle forming at the three phase contact line, from the height of
the jump in the droplet center distance. If $\tilde{R}$ is the
radius of the droplet after the membrane has formed, i.e., the
radius of an ideal sphere fitted to the free droplet surface, it is
easily shown that the volume of the flattened droplet is given by
\begin{equation}
V(\theta) = \frac{\pi \tilde{R}}{3}\left(2+3\cos\frac{\theta}{2}-\cos^3\frac{\theta}{2}\right)
\end{equation}
where $\theta$ is defined as above. If $D$ is the distance of the
sphere centers after membrane formation, we furthermore have
$D=2\tilde{R}\cos\theta$. Denoting by $R$ the radius of the droplets
before membrane formation, we finally obtain using volume
conservation
\begin{equation}
\frac{2}{\cos^3\theta}+\frac{3}{\cos^2\theta} = 1+4\left(\frac{R}{D}\right)^3
\end{equation}
From the height of the jump in the black curve in
Fig.~\ref{MembraneCoupling}, we obtain $2R/D \approx 1.107$ and thus
$\theta \approx 51$ degrees, in reasonable agreement with the value
found above.

Here we are particularly interested in the change in coupling
induced by the membrane formation. Clearly, we see that after the
jump in the black curve, all three transmittance traces are firmly
coupled. It seems that due to the presence of the mono-olein double
bonds in the oil phase, an efficient transport of bromine as a
messenger through the oil phase is prevented effectively. Transport
through the mono-olein bilayer membrane, however, is obviously still
possible.

\begin{figure}[h]
\includegraphics[width = 8.5cm]{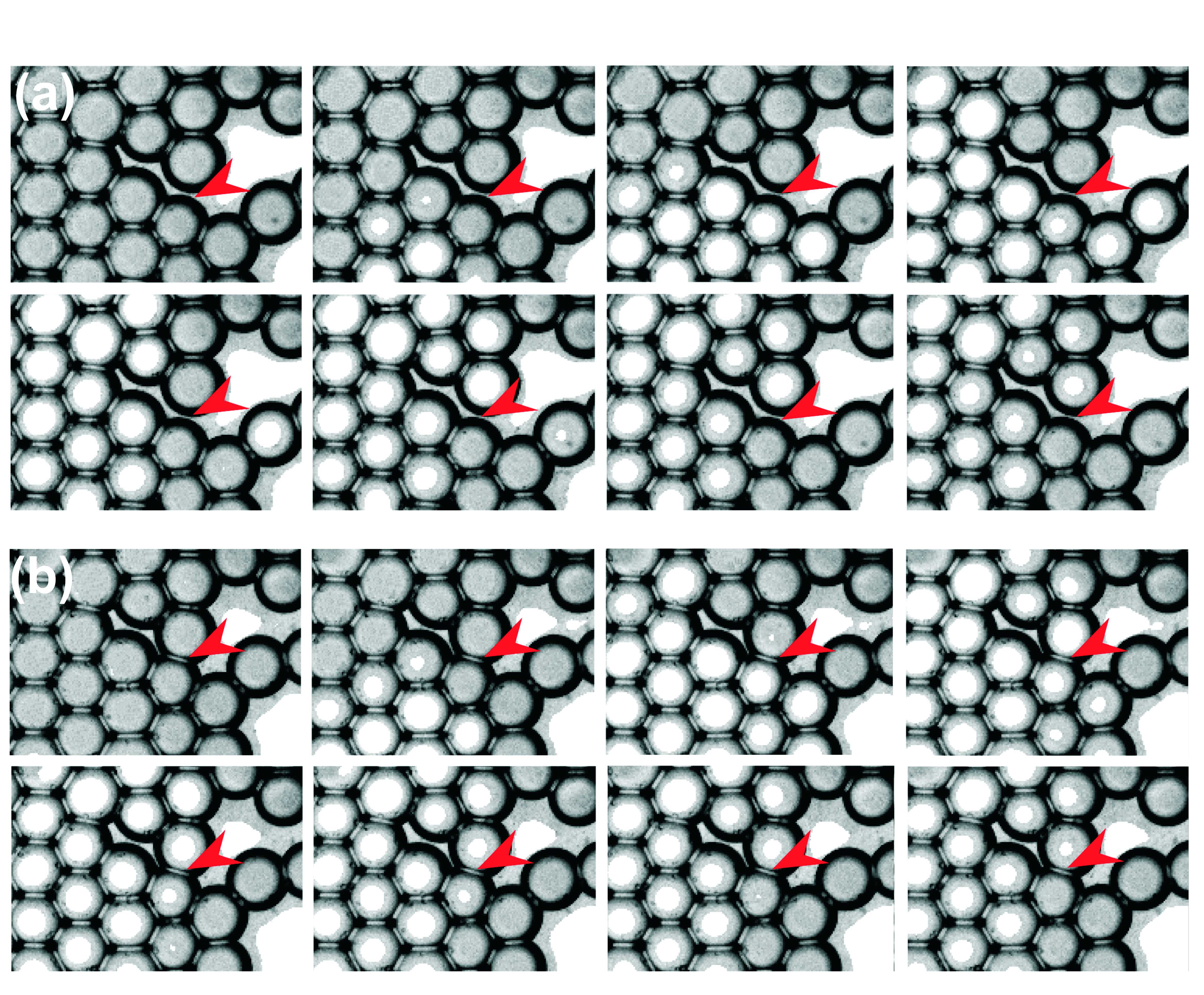}
\caption{Demonstration of the influence of membrane coupling on an excitation wave in a larger raft of droplets. The oil phase consists of squalane with 50 mM/l mono-olein. The aqueous phase is 0.5 M sulphuric acid, 280 mM sodium bromate, 0.5 M malonic acid, and 3 mM ferroin. (a) The droplet surfaces indicated by the red arrow are very close, but there has yet no membrane been formed. The excitation wave (coming from below) passes around the arrow, the excitation does not directly cross the gap between the two droplets. (b) After the membrane has formed, the excitation wave passes through the site indicated by the arrow.\label{RaftWave} }
\end{figure}

The strong variation in coupling strength by membrane formation can be seen most impressively in excitatory waves propagating in rafts of BZ droplets with a yet incomplete network of bilayer membrane contacts. Fig.~\ref{RaftWave} shows snapshots of such a wave of excitation (i.e., of a sudden increase in transmittance) which propagates from bottom to top in a moderately dense raft of droplets. The top and bottom row each shows a time lapse representation of a single excitation wave. The red arrow points to a droplet/droplet contact which in the top row has not yet formed a membrane, but in the bottom row it has. A close inspection shows that in the top row, the wave travels around the point of the red arrow. In the bottom row, however, it passes this contact without noticeable hesitation.

\section{conclusions}

Self assembled surfactant bilayer networks in microfluidic channels may be provide a crucial first step towards complex dynamical functions comprising nanoscale or molecular units. More specifically, native surfactant bilayers already offer a range of different electrical behaviour that can be exploited to create wet circuitry. The stability of these objects in micro-fluidic systems is quite encouraging, both in static and in dynamic settings. Their employment as externally controlled scaffolds for synthetic functional molecular units thus appears feasible. The peculiar permeation properties of bilayer membranes for messenger molecules, such as occur in systems of chemical oscillators, furthermore suggests the development of multi-functional, self-assembling dynamic nanoscale systems which open up novel types of soft matter technology, which is conceptually influenced by the physical building principles of living matter, but relies on simple components apt to synthesis and thorough control.

Clearly, there is still a long way to go before emulsion-based self-assembled functional systems superior to conventional top-down devices can be produced. This does not only concern the problem of length scales (which appears favorably addressable, as discussed above), but also the question of how to direct many different droplet contents and membrane compounds into the desired patterns. First steps have been undertaken, but the majority of an exciting pathway is still ahead.

\section*{Acknowledgements}

This work has been supported by the Deutsche Forschungsgemeinschaft within the Collaborative Research Center on 'Nanoscale Photonic Imaging' (SFB 755).

\end{document}